\documentclass[aps,amsfonts,pra,twocolumn,showpacs,superscriptaddress]{revtex4-1}
\usepackage{epsfig,amsmath,amssymb,bm,epsf,graphicx,psfrag,float}
\usepackage{diagbox}
\usepackage{slashbox}

\usepackage{amsfonts}

\usepackage{color}
\usepackage{xcolor}
\usepackage{mathtools}

\usepackage{bbm}
\usepackage{enumerate}
\usepackage[inline,final]{showlabels}

\usepackage[all]{xy}
\usepackage{color}
\usepackage[utf8]{inputenc}
\usepackage[T1]{fontenc}
\usepackage{lmodern, mathtools}
\newtheorem{prop}{Proposition}

\def\bra#1{\langle#1\vert}
\def\ket#1{\vert#1\rangle}

\DeclareFontFamily{U}  {MnSymbolA}{}
\DeclareSymbolFont{MnSyA}         {U}  {MnSymbolA}{m}{n}
\SetSymbolFont{MnSyA}       {bold}{U}  {MnSymbolA}{b}{n}
\DeclareFontShape{U}{MnSymbolA}{m}{n}{
<-6>  MnSymbolA5
<6-7>  MnSymbolA6
<7-8>  MnSymbolA7
<8-9>  MnSymbolA8
<9-10> MnSymbolA9
<10-12> MnSymbolA10
<12->   MnSymbolA12}{}
\DeclareFontShape{U}{MnSymbolA}{b}{n}{
<-6>  MnSymbolA-Bold5
<6-7>  MnSymbolA-Bold6
<7-8>  MnSymbolA-Bold7
<8-9>  MnSymbolA-Bold8
<9-10> MnSymbolA-Bold9
<10-12> MnSymbolA-Bold10
<12->   MnSymbolA-Bold12}{}

\makeatletter
\newcommand{\xleftfork}[2][]{%
\ext@arrow 0079\xleftforkfill@{#1}{#2}%
}
\newcommand{\xleftforkfill@}{%
\arrowfill@\Mnrelbar\Mnrelbar{\mathrel{\leftpitchfork}}
}
\newcommand{\xrightfork}[2][]{%
\ext@arrow 0097\xrightforkfill@{#1}{#2}%
}

\DeclareMathSymbol{\rightpitchfork}{\mathrel}{MnSyA}{"88}%
\DeclareMathSymbol{\leftpitchfork}{\mathrel}{MnSyA}{"8A}%
\DeclareMathSymbol{\Mnrelbar}{\mathrel}{MnSyA}{"D0}%

\def\bra#1{\langle#1\vert}
\def\ket#1{\vert#1\rangle}

\def\ipr#1#2{\langle#1\vert#2\rangle}

\newcommand{\revb}[1]{{\color{violet}#1}}

\newcommand{\rpf}{\rightpitchfork}

\newcommand{\lpf}{\leftpitchfork}

\begin{document}

\title{AKLT models on decorated square lattices are gapped}

\author{Nicholas Pomata}
\affiliation{C. N. Yang Institute for Theoretical Physics and Department of Physics and Astronomy, State University of New York at Stony Brook, Stony Brook, NY 11794-3840, USA}
\author{Tzu-Chieh Wei}
\affiliation{C. N. Yang Institute for Theoretical Physics and Department of Physics and Astronomy, State University of New York at Stony Brook, Stony Brook, NY 11794-3840, USA}
\affiliation{Institute for Advanced Compuational Science, State University of New York at Stony Brook, Stony Brook, NY 11794-5250, USA}
\date{\today}

\begin{abstract}
 The nonzero spectral gap of the original two-dimensional Affleck-Kennedy-Lieb-Tasaki (AKLT) models has remained unproven for more than three decades. Recently, Abdul-Rahman et al. [arXiv:1901.09297] provided an elegant approach and proved analytically the existence of a nonzero spectral gap for the AKLT models on the decorated honeycomb lattice (for the number $n$ of spin-1 decorated sites on each original edge no less than 3). We perform calculations for the decorated square lattice and show that the corresponding AKLT models are gapped if $n\ge 4$.  Combining both results, we also show that a family of decorated hybrid AKLT models, whose underlying lattice is of mixed vertex degrees 3 and 4, are also gapped for $n\ge 4$. We develop a numerical approach that extends beyond what was accessible previously. Our numerical results further improve the nonzero gap to $n\ge 2$, including the establishment of the gap for $n=2$ in the decorated triangular and cubic lattices.  The latter case is interesting, as this shows the AKLT states on the decorated cubic lattices are not N\'eel ordered, in contrast to the state on the undecorated cubic lattice.
\end{abstract}

 \maketitle

 \section{Introduction}
 Affleck, Kennedy, Lieb, and Tasaki (AKLT) constructed a  one-dimensional spin-1 chain whose Hamiltonian is rotation-invariant in the spin degree of freedom~\cite{AKLT1}, but has a spectral gap above the unique ground state, in contrast to the spin-1/2 antiferromagnetic Heisenberg model. This provided strong support for  Haldane's conjecture~\cite{Haldane83,Haldane83b} regarding the relation between the spectral gap and spin magnitudes in quantum magnetism. They also generalized the construction to two dimensions~\cite{AKLT2}, and showed, in particular, that the spin-spin correlation function of the ground-state wavefunction decays exponentially in the honeycomb and the square lattice models. The uniqueness of the ground state in these models was further analyzed by Kennedy, Lieb and Tasaski~\cite{KLT}.  There have been a few useful techniques for showing uniqueness of the ground state and gap~\cite{Fannes1992,KLT,Knabe}, which work well in one dimension, but the proof of the nonzero spectral gap has not been established for either of the two 2D AKLT models, even more than three decades after their construction.

 Haldane's conjecture on the spectral property of isotropic chains of integer spins complements the result of Lieb, Schultz and Mattis (LSM) on the properties of chains of half-odd spins, which states that there exists an excited state with energy degenerate with the ground state in the thermodynamic limit~\cite{LSM}.   That is the system is either gapless or has degenerate ground states. This LSM theorem was generalized to   higher dimensions~\cite{Oshikawa,Hastings} with each unit cell having half-integer total spin, and the ground state, in addition to the possibility of being gapless or degenerate, can also be a gapped spin liquid that does not break the symmetry.
 Recently, due to the tremendous progress on topological phases, the LSM theorem has been re-examined in new perspectives, such as symmetry-protected topological (SPT) phases,  crystalline symmetry, anomaly, and boundary~\cite{PoWatanabe17,JianBiXu,Lu,Cheng,Shiozaki,OgataTasaki,ChoHsiehRyu,YaoOshikawa}.  For example, it was conjectured that all LSM-like
theorems can be understood from lattice homotopy~\cite{PoWatanabe17}, and this was very recently generalized  to develop a topological theory of LSM theorems in quantum spin systems~\cite{ElseThorngren}.

 Unexpectedly, 2D AKLT states have recently emerged as resource for univeral quantum computation (QC) in the framework of the measurement-based quantum computation (MBQC)~\cite{RB2001,BBDR2009,RW2012,Wei2018}.  The spin-3/2 AKLT state on the honeycomb  lattice was first shown to provide the appropriate entanglement structure for universal QC~\cite{Wei2011,Miyake2011}, a result subsequently generalized to  other trivalent lattices~\cite{Wei2013}. Before the demonstration of the computation\revb{al} universality of the spin-2 AKLT state on the square lattice~\cite{Wei2015}, a few decorated lattice structures (with mixed vertex degrees) and the corresponding AKLT states were first considered in Ref.~\cite{Wei2014}. A {partial} picture of quantum computation universality in the family of  AKLT states is as follows.  Any AKLT state 
{residing} on a two- or three-dimensional frustration-free regular lattice (no loop with {an} odd number of sites) with any combination of spin-2,
spin-3/2, spin-1, and spin-1/2  that is consistent with the lattice. Higher-spin systems are mainly not included due to technicalities~\cite{Wei2015}.

 Regarding  the gap,   tensor network methods were employed and the value of the gap in the thermodynamic limit was estimated~\cite{Garcia-Saez2013,Vanderstraeten2015}.  A recent breakthrough in the analytic proof was given by Abdul-Rahman et al.~\cite{Abdul-Rahman2019}, who considered a family of  decorated honeycomb lattices and proved that the corresonding AKLT models are gapped for the number $n$ of decorated sites being greater than 2; see e.g. Fig.~\ref{fig:lattices}a. The associated AKLT states, according to the results of  Ref.~\cite{Wei2014}, are also universal for MBQC, and hence are also of  interest, as the non-zero gap implies that preparation of these states via cooling is useful.   Additional progress in analytics has also been made by Lemm, Sandvik and Yang on hexagonal chains~\cite{Lemm2019}, where the quasi-1D AKLT models are also gapped.
 
 We note that the results of Ref.~\cite{Abdul-Rahman2019}, as argued below, apply directly to  other trivalent lattices with decoration, such as the square-octagon $(4,8^2)$, the cross $(4,6,12)$, and the star $(3,12^2)$ (Fig.~\ref{fig:lattices}b,c\&d.)  Although the AKLT Hamiltonians are frustration-free, some features in generalized measurement display some frustration, e.g. on the star lattice~\cite{Wei2013}. The decoration renders the frustrated star lattice non-frustrated and removes the frustration features in the measurement.  AKLT states on all these decorated lattices are also universal for MBQC~\cite{Wei2013,Wei2014}.

  Here we prove analytically that AKLT models on 2D decorated square lattices possess nonzero spectral gap for $n\ge 4$, where $n$ is the number of spin-1 decorated sites added to each original edge (see e.g. Fig.~\ref{fig:lattices}e\&f). This result also implies, in addition to the decorated kagome and $(3,4,6,4)$ lattices (Fig.~\ref{fig:lattices}g\&h), decorated 3D diamond lattices host AKLT models with nonzero spectral gap.  AKLT states on the 3D diamond lattice and the associated decorated ones are also universal~\cite{Wei2014,Wei2015}, and the significance is that these 3D resource states are likely to provide fault-tolerance similar to the 3D cluster state~\cite{RHG}.
 Moreover, proving the spectral gap and knowing its value will be crucial in state preparation and validation protocols. 
 
 Using the results from both the decorated honeycomb and square lattice, we also show that AKLT models on decorated lattices whose underlying lattice is of mixed vertex degrees 3 and 4 are also gapped for $n\ge 4$. We also provide a numerical approach that allows us to study the parameters which bound the gap for $n>1$, previously thought inaccessible. Our numerical results further improve the nonzero gap to $n\ge 2$, including the establishment of the gap for $n=2$ in the decorated triangular and cubic lattices, i.e. those whose underlying lattices \revb{have} vertex degree 6.   We also provide much improved lower bounds on the spectral gap for some of the AKLT models. The structure of the remaining paper is as follows.
  In Sec.~\ref{sec:review} we first review methods used in Ref.~\cite{Abdul-Rahman2019}. Then in Sec.~\ref{sec:calculations} we perform the same detailed calculations for the AKLT models on the decorated square lattices. In Sec.~\ref{sec:honeycomb} we make some comments on the other decorated lattices. In Sec.~\ref{sec:numerics}, we describe our numerical methods {which} improve all {the} above gappedness scenarios to $n\ge2$. Finally in Sec.~\ref{sec:conclude} we make some concluding remarks.
 
 \section{Review of prior methods and results}
 \label{sec:review}
 
 \begin{figure}
 {\centering
 \includegraphics[width=0.45\textwidth]{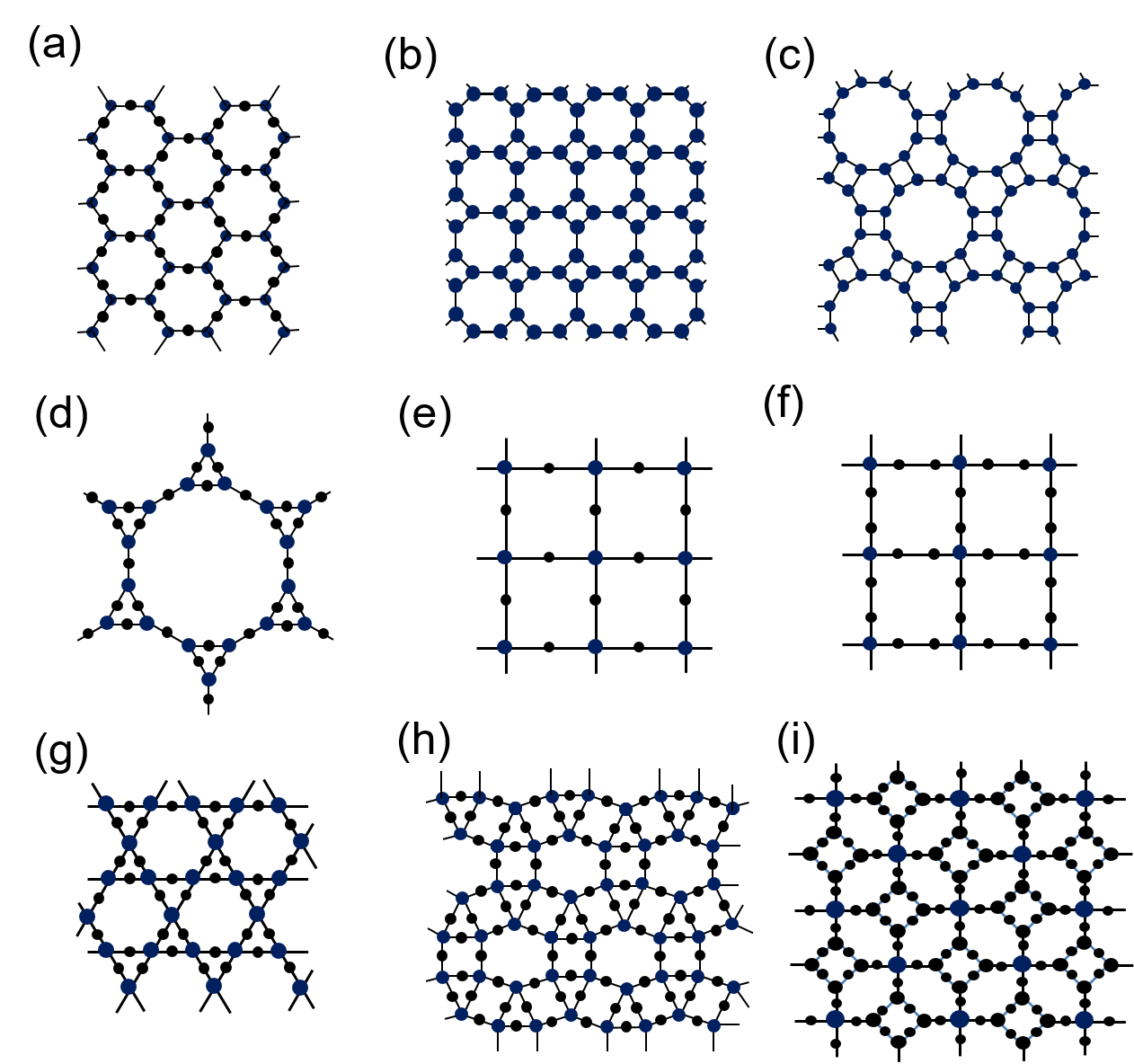}}
 \caption{\label{fig:lattices}
 Illustration of lattices. Some are decorated: (a), (d), (e), (f), (g) \& (i), namely, with additional sites added to underlying lattices. Underlying lattices of (a)-(d) are trivalent; underlying lattices of (e)-(h) are four-valent; the underlying lattice of (i) is of mixed vertex degrees of 3 and 4. }
 \end{figure}
 
 Here we briefly review the key points that enable the proof of the spectral gap for AKLT models on the decorated honeycomb lattice in Ref.~\cite{Abdul-Rahman2019}; see Fig.~\ref{fig:lattices} for one such illustration  with $n=1$, as well as other lattices. We will try to use the same symbols \revb{as} in Ref.~\cite{Abdul-Rahman2019} as much as possible, but may have some slight differences. Consider an original lattice $\Lambda$ (e.g. honeycomb or square lattice) and its decorated version $\Lambda^{(n)}$ in which each edge of $\Lambda$ has been decorated with $n$ spin-1 sites. 
  Let ${\cal E}_{\Lambda^{(n)}}$ denote the edge set of the decorated lattice.  
 The AKLT model Hamiltonian defined on $\Lambda^{(n)}$ is
 \begin{equation}
 H_{\Lambda^{(n)}}^{\rm AKLT}=\sum_{e\in {\cal E}_{\Lambda^{(n)}} }P_e^{(z(e)/2)},
 \end{equation}  
 where $P_e^{(z(e)/2)}$ is a projection onto the total spin $s=z(e)/2$ subspace of the two spins linked by the edge $e$, and $z(e)$ denotes the sum of the coordination numbers (i.e. vertex degrees $z_a$ and $z_b$) of the two spins $a$ and $b$ linked by edge $e$. 
 
\begin{figure}
 {\centering
 \includegraphics[width=0.45\textwidth]{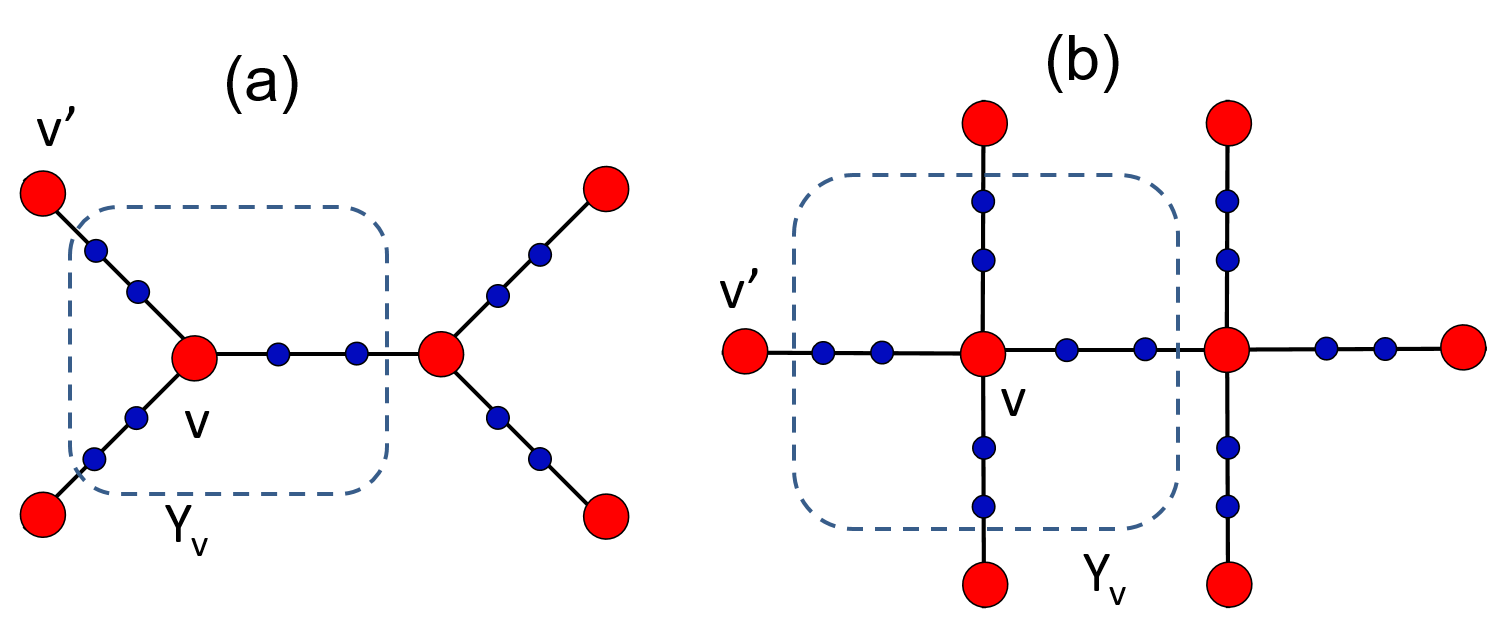}}
 \caption{\label{fig:Yv}
 Illustration of local structure of decorated lattices: (a) the decorated honeycomb and (b) the decorated square lattice, both with $n=2$. }
 \end{figure}

 Instead of directly using the AKLT Hamiltonian, Ref.~\cite{Abdul-Rahman2019} first considers a slightly modified one: 
 \begin{equation}
 \label{eqn:HY}
 H_{Y}\equiv \sum_{v\in\Lambda} h_v=\sum_{v\in\Lambda} \sum_{e\in {\cal E}_{Y_v}} P_e^{(z(e)/2)},
 \end{equation}
where  $h_v$ is the AKLT Hamiltonian on the set $Y_v$ of $(zn+1)$ vertices of the decorated lattice $\Lambda^{(n)}$, $z$ is the coordination number ($z=3$ for the honeycomb) and ${\cal E}_{Y_v}$ denotes the edges connecting vertices in $Y_v$; see Fig.~\ref{fig:Yv} for illustration. It has a few terms in $H_{\Lambda^{(n)}}^{\rm AKLT}$ missing, i.e., those terms on the edges containing the last spin-1 site on edge $e\in {Y_v}$ and the next site $v' \in \Lambda$. So we have an inequality
\begin{equation}
H_{\Lambda^{(n)}}^{\rm AKLT}\le H_{Y}\le 2H_{\Lambda^{(n)}}^{\rm AKLT}.
\end{equation}
 However, instead of $H_Y$, Ref.~\cite{Abdul-Rahman2019} also considers a slight modification
 \begin{equation}
 \tilde{H}_{\Lambda^{(n)}}\equiv \sum_{v\in\Lambda} P_v,
 \end{equation}
 where $P_v$ is the orthogonal projection onto the range of $h_v$.  The kernel of $P_v$ is the ground space of $h_v$, i.e., ${\rm ker} P_v={\rm ker} h_v$. Then it is shown that 
 \begin{equation}
 \label{eqn:upperlowerbounds}
 \frac{\gamma_Y}{2} \tilde{H}_{\Lambda^{(n)}} \le H_{\Lambda^{(n)}}^{\rm AKLT}\le \|h_{v}\| \tilde{H}_{\Lambda^{(n)}},
 \end{equation}
 where $\gamma_Y$ is the smallest nonzero eigenvalue of $h_v$ (or equivalently the spectral gap of the small system $Y_v$) and $\|h_{v}\|$ is the usual operator norm of $h_{v}$ (or equivalently the largest eigenvalue of $h_v$, since $h_v$ is non-negative).

 The strategy is to prove $\tilde{H}_{\Lambda^{(n)}}$ is gapped. By squaring $\tilde{H}_{\Lambda^{(n)}}$, we find that
 \begin{eqnarray}
 (\tilde{H}_{\Lambda^{(n)}})^2&=&\tilde{H}_{\Lambda^{(n)}}+\sum_{v\ne w} (P_v P_w + P_w P_v)\\
 &\ge& \tilde{H}_{\Lambda^{(n)}}+\sum_{(v,w)\in {\cal E}_\Lambda} (P_v P_w + P_w P_v),
 \end{eqnarray}
 where  for those $v$ and $w$ not on the same edge $P_v P_w$ is non-negative and is dropped, resulting in the last inequality.
 If one can find the minimum positive number $\eta>0$ such that $P_v P_w + P_w P_v \ge -\eta (P_v+P_w)$, then
  \begin{eqnarray}
 (\tilde{H}_{\Lambda^{(n)}})^2&\ge &\tilde{H}_{\Lambda^{(n)}}-\eta_n\sum_{(v,w)\in {\cal E}_\Lambda}(P_v + P_w )\\
 &=& (1-z\eta_n)\tilde{H}_{\Lambda^{(n)}}= \gamma\tilde{H}_{\Lambda^{(n)}},
 \label{eqn:ham-ineq}
 \end{eqnarray}
 where $\gamma\equiv 1-z\eta_n$ (the subscript $n$ is added to $\eta$) and $z$ is the coordination number of the underlying lattice $\Lambda$ (e.g. $z=3$ for the honeycomb and $z=4$ for the square lattice).
 If $\gamma>0$, then one proves that $\tilde{H}_{\Lambda^{(n)}}$ has a spectral gap above the ground state(s).
 
 Therefore, most of effort goes into finding $\eta$ or an upper bound. A relation that was used to this end in Ref.~\cite{Abdul-Rahman2019} is Lemma 6.3 from Ref.~\cite{Fannes1992} for a pair of projectors $E$ and $F$:
 \begin{equation}
 EF+FE\ge -\|EF-E\wedge F\| (E+F),
 \end{equation}
 where $E\wedge F$ denotes the projection onto the joint subspace $ E\mathcal{H} \cap  F\mathcal{H}$.  When we apply this relation to \eqref{eqn:ham-ineq},  $\varepsilon_n=\|EF-E\wedge F\|$ becomes an upper bound on $\eta_n$, i.e., $\eta_n\le \varepsilon_n$. In particular, in Prop.~\ref{prop:EplusF} below, we determine that $\eta_n=\varepsilon_n$. In Sec.~\ref{sec:numerics} below we will additionally develop techniques to compute $\eta_n$ exactly.
 
  \begin{figure}
 {\centering
 \includegraphics[width=0.48\textwidth]{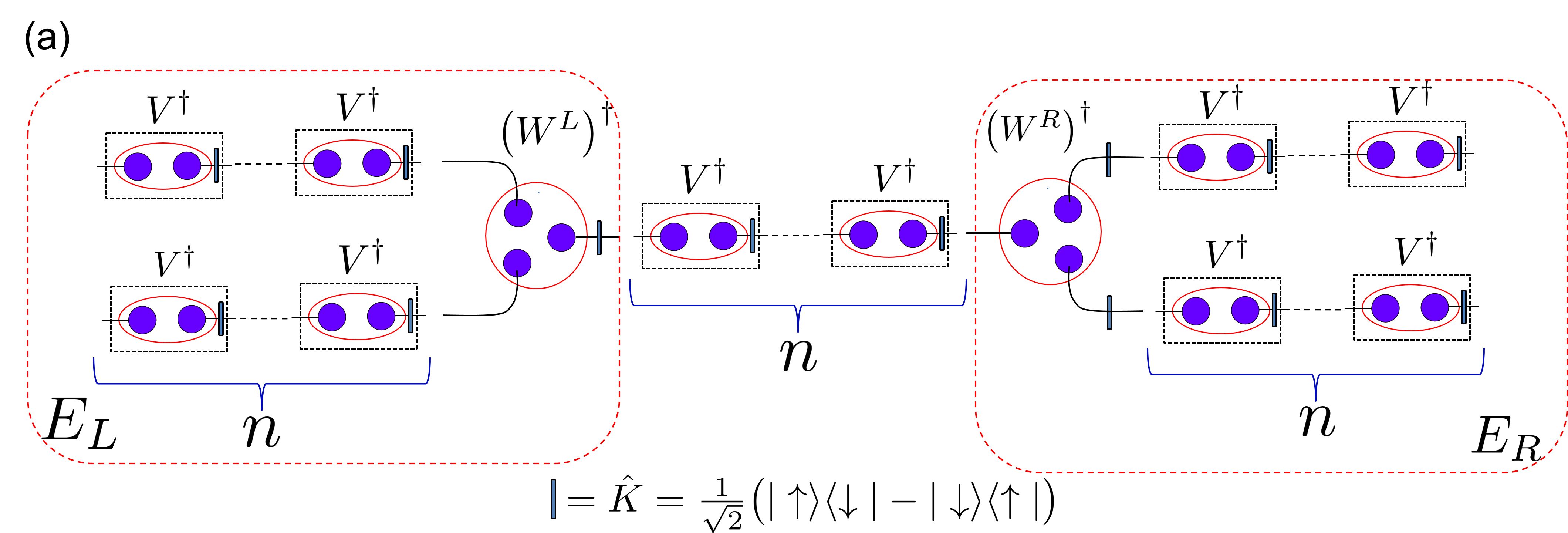}
 \includegraphics[width=0.48\textwidth]{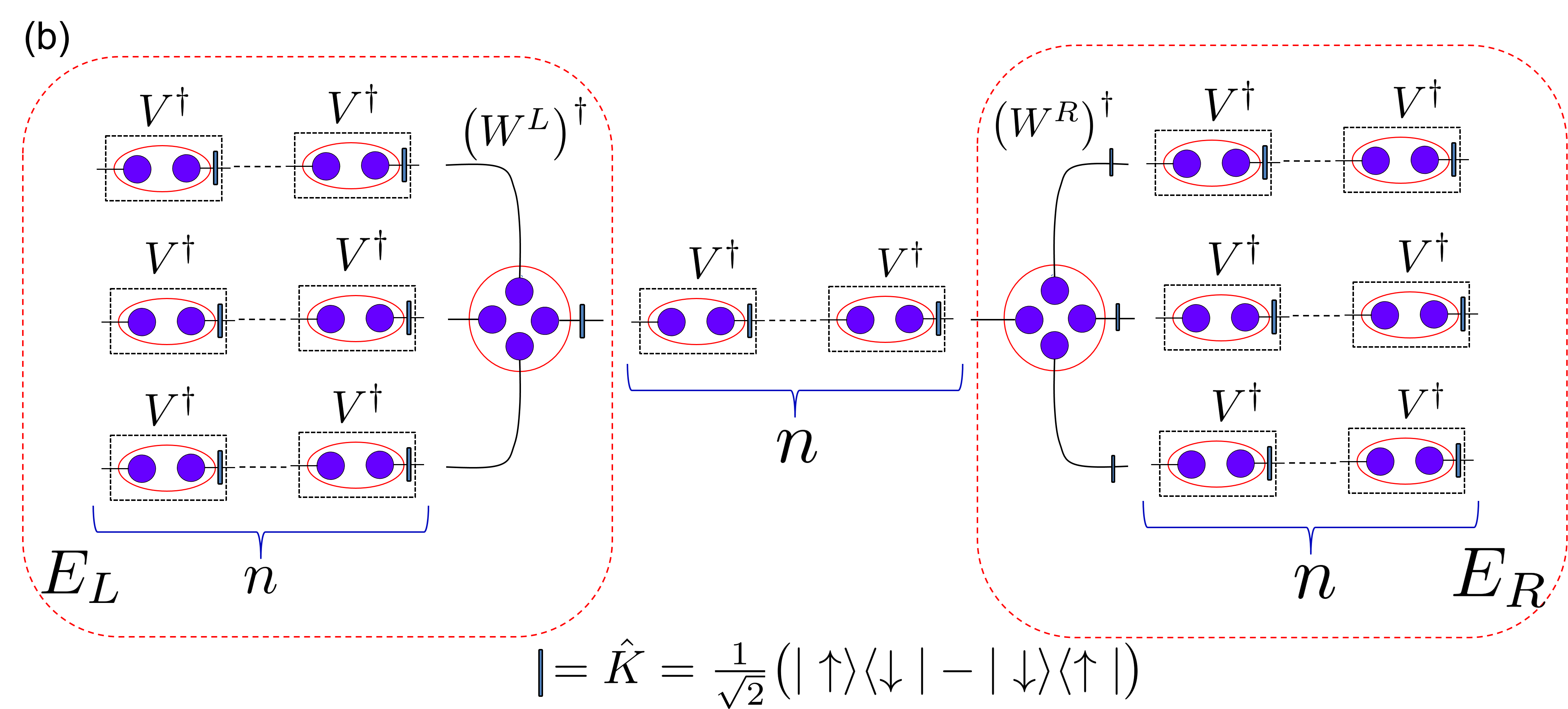}}
 \caption{\label{fig:tensors}
 Illustration of local lattice structure and tensors for (a) the honeycomb or any trivalent lattices and (b) square lattice or any four-valent lattices. The solid purple dots represent virtual qubits. The solid line between two neighboring qubits represent a maximally entangled state of the form $|\phi^+\rangle\equiv(|\uparrow\uparrow\rangle+|\downarrow\downarrow\rangle)/\sqrt{2}$. The solid vertical bar denotes the operator $\hat{K}=(|\uparrow\rangle\langle\downarrow|-|\downarrow\rangle\langle\uparrow|)/\sqrt{2}$, which map $|\phi^+\rangle$ to a singlet, up to normalization. The tensors $T_l^L$ consist of $W^L$ and $V$'s inside the dotted square labeled by the channel's symbol $E_L$, and similarly the tensors $T_r^R$ consist of $W^R$ and $V$'s inside the dotted square labeled by $E_R$.}
 \end{figure}
 
 Using the above Lemma and employing tensor-network approaches, the authors of Ref.~\cite{Abdul-Rahman2019} show elegantly that 
 \begin{equation}
 \label{eqn:bound}
\varepsilon_n \le \frac{4\cdot 3^{-n}}{\sqrt{1-b_{LR}(n)} }+ \left(\frac{16\cdot 3^{-2n}}{1-b_{LR}(n)}\right) \big(1+ b_G (n)\big),
 \end{equation} 
 where 
 \begin{eqnarray}
 b_G(n)&\equiv&\frac{4\cdot 3^{-n}}{q_L(n)q_R(n)}\|E_L\|\,\|E_R\|,\\
 b_L(n)&\equiv&\frac{8\cdot 3^{-n}}{q_L(n)}\|E_L\|,\\
 b_R(n)&\equiv&\frac{4\cdot 3^{-n}}{q_R(n)}\|E_R\|,\\
 b_{LR}(n)&\equiv&b_L(n)+b_R(n)-b_L(n)\,b_R(n).
 \end{eqnarray}
 In the above expressions, $E_L$ is  the quantum channel, or equivalently the transfer matrix, obtained from  the tensors $T^L$ associated with the `left' set of vertices $Y_v\setminus Y_w$
and $E_R$  (via tensors $T^R$) is associated with the `right' set of vertices $Y_w\setminus Y_v$.  See also Figs.~\ref{fig:Yv} and~\ref{fig:tensors} for illustration. More precisely, the channels are defined as follows,
 \begin{equation}
 E_L(B)=\sum_l (T_l^L)^\dagger B T_l^L, \ \ E_R(C) =\sum_r T_r^R C (T_r^R)^\dagger.
 \end{equation}
Note that by examining the derivations in Ref.~\cite{Abdul-Rahman2019}, the operator norms associated with $\|E_L\|$ and $\|E_R\|$  holds both for the norm with respect to $C^*$-norm of $B$ and $C$ and for that w.r.t. the Hilbert-Schmidt norm of matrices.
However, since the latter norm is larger for the former, the former norm presents a better bound. 

  Moreover, two specific matrices are introduced: $Q_L\equiv E_L(\openone)$ and $Q_R\equiv E_R^t(\rho_1)$ ($\rho_1$ here equaling $\mathbbm{1}/2$), and $q_L$ and $q_R$ are their respective minimum eigenvalues. For the proof, we highly recommend Ref.~\cite{Abdul-Rahman2019} to the readers.

 \section{Analysis of spectral gap}
 \label{sec:calculations}
 The spin-2 entity residing on each square lattice site is composed of four virtual qubits projected onto their symmetric subspace, and the mapping between the physical spin-2 degrees of freedom and the those in the symmetric subspace is as follows,
 \begin{eqnarray}
P_{\rm sym}=&&|2\rangle\langle\uparrow\uparrow\uparrow\uparrow\!|+|-2\rangle\langle\downarrow\downarrow\downarrow\downarrow\!|
 \nonumber\\
 &+&
 |1\rangle\frac{1}{2}(\langle\downarrow\uparrow\uparrow\uparrow\!|+\langle\uparrow\downarrow\uparrow\uparrow\!|+\langle\uparrow\uparrow\downarrow\uparrow\!|+\langle\uparrow\uparrow\uparrow\downarrow\!|)\nonumber\\
 &+&|-1\rangle\frac{1}{2}(\langle\uparrow\downarrow\downarrow\downarrow\!|+\langle\downarrow\uparrow\downarrow\downarrow\!|+\langle\downarrow\downarrow\uparrow\downarrow\!|+\langle\downarrow\downarrow\downarrow\uparrow\!|)\nonumber\\
 &+&|0\rangle\frac{1}{\sqrt{6}}(\langle\uparrow\uparrow\downarrow\downarrow\!|+\langle\uparrow\downarrow\uparrow\downarrow\!|+\langle\downarrow\uparrow\uparrow\downarrow\!|+\langle \uparrow\downarrow\downarrow\uparrow\!|\nonumber\\
 &&+\,\langle\downarrow\uparrow\downarrow\uparrow\!|+\langle\downarrow\downarrow\uparrow\uparrow\!|),\nonumber
 \end{eqnarray} 
 where $|m\rangle$'s are eigenstates of spin-2 $S_z$ operators with eigenvalue $m$'s. If we consider one square lattice site on the left, then there are corresponding tensors for $P_m^L$, which are
 \begin{eqnarray}
  &&P_{2}=|\!\uparrow\rangle\langle\uparrow\uparrow\uparrow\!|,\ P_{-2}=|\!\downarrow\rangle\langle\downarrow\downarrow\downarrow\!|
 \nonumber\\
 && P_1=\frac{1}{2}|\!\downarrow\rangle\langle\uparrow\uparrow\uparrow\!|+\frac{1}{2}|\!\uparrow\rangle(\langle\downarrow\uparrow\uparrow\!|+\langle\uparrow\downarrow\uparrow\!|+\langle\uparrow\uparrow\downarrow\!|),\nonumber\\
 &&P_{-1}=\frac{1}{2}|\!\uparrow\rangle\langle\downarrow\downarrow\downarrow\!|+\frac{1}{2}|\!\downarrow\rangle(\langle\uparrow\downarrow\downarrow\!|+\langle\downarrow\uparrow\downarrow\!|+\langle\downarrow\downarrow\uparrow\!|),\nonumber\\
 &&P_0=\frac{1}{\sqrt{6}}|\!\uparrow\rangle(\langle\uparrow\downarrow\downarrow\!|+\langle\downarrow\uparrow\downarrow\!|+\langle \downarrow\downarrow\uparrow\!|),\nonumber\\
 & & \qquad +\frac{1}{\sqrt{6}}|\!\downarrow\rangle(\langle\uparrow\uparrow\downarrow\!|+\langle\uparrow\downarrow\uparrow\!|+\langle\downarrow\uparrow\uparrow\!|).\nonumber
 \end{eqnarray} 
Because the AKLT state is formed from projecting virtual singlet pairs via symmetric projectors, we obtain the local tensors describing the spin-2 site on the left as $W_k^L\equiv \sqrt{2} K P_k$, where $K= (|\!\uparrow\rangle\langle\downarrow\!|-|\!\downarrow\rangle\langle\uparrow\!|)/\sqrt{2}$, and they are given as follows,
\begin{eqnarray}
  &&W^L_{2}=-|\!\downarrow\rangle\langle\uparrow\uparrow\uparrow\!|, \ \ W^L_{-2}=|\!\uparrow\rangle\langle\downarrow\downarrow\downarrow\!|,
 \nonumber\\
 && W^L_1=\frac{1}{2}|\!\uparrow\rangle\langle\uparrow\uparrow\uparrow\!|-\frac{1}{2}|\!\downarrow\rangle(\langle\downarrow\uparrow\uparrow\!|+\langle\uparrow\downarrow\uparrow\!|+\langle\uparrow\uparrow\downarrow\!|),\nonumber\\
 &&W^L_{-1}=-\frac{1}{2}|\!\downarrow\rangle\langle\downarrow\downarrow\downarrow\!|+\frac{1}{2}|\!\uparrow\rangle(\langle\uparrow\downarrow\downarrow\!|+\langle\downarrow\uparrow\downarrow\!|+\langle\downarrow\downarrow\uparrow\!|),\nonumber\\
 &&W^L_0=-\frac{1}{\sqrt{6}}|\!\downarrow\rangle(\langle\uparrow\downarrow\downarrow\!|+\langle\downarrow\uparrow\downarrow\!|+\langle \downarrow\downarrow\uparrow\!|),\nonumber\\
 & & \qquad\quad +\frac{1}{\sqrt{6}}|\!\uparrow\rangle(\langle\uparrow\uparrow\downarrow\!|+\langle\uparrow\downarrow\uparrow\!|+\langle\downarrow\uparrow\uparrow\!|).\nonumber
 \end{eqnarray} 
 See also Fig.~\ref{fig:tensors}b for illustration of the local lattice structure and the corresponding tensors.
 From these, one can easily check that
 \begin{equation}
 \label{eqn:WW}
 \sum_k W_k^L (W_k^L)^\dagger=\frac{5}{2}\openone_{C^2},
 \end{equation}
 and one can define a quantum channel 
 \begin{equation}
 \label{eqn:Erpf}
 E^\rpf(B)\equiv \sum_i (W_i^L)^\dagger B W_i^L.
 \end{equation} 
 (We note that one could re-scale $W^L$'s so as to make the right hand side of Eq.~(\ref{eqn:WW}) be $\openone$, but we won't do that here.) One also finds that
 \begin{equation}
 \label{eqn:EI}
 E^\rpf(\openone)=\frac{5}{4} \Pi^{S=3/2}_{\rm sym},
 \end{equation} 
 where $\Pi^{S=3/2}_{\rm sym}$ is the projector to the 3-qubit symmetric subspace. 
 At this point it is useful to introduce the two W states used in quantum information so as to simplify the notation,
 \begin{eqnarray}
 |w\rangle&\equiv&\frac{1}{\sqrt{3}}(|\uparrow\uparrow\downarrow\rangle+|\uparrow\downarrow\uparrow\rangle+|\downarrow\uparrow\uparrow\rangle),\\
   |\tilde{w}\rangle&\equiv&\frac{1}{\sqrt{3}}(|\uparrow\downarrow\downarrow\rangle+|\downarrow\uparrow\downarrow\rangle+| \downarrow\downarrow\uparrow\rangle).
 \end{eqnarray}
 The associated dual quantum channel is defined as 
 ${E^\rpf}^t(B)\equiv \sum_i W_i^L B (W_i^L)^\dagger$, which maps any three-qubit density matrix to a one-qubit density matrix, and can be written as (assuming $B$ is Hermitian for simplicity)
 \begin{equation}
 {E^\rpf}^t(B) =c_0(B) \openone + c_x (B) \sigma^x + c_y(B) \sigma^y+ c_z (B)\sigma^z,
 \end{equation}
 where the four coefficients $c\revb{_i}$ are
 \begin{eqnarray}
 c_0(B)&=&\frac{5}{8}(\langle\uparrow\uparrow\uparrow\!|B|\!\uparrow\uparrow\uparrow\rangle+\langle\downarrow\downarrow\downarrow\!|B|\!\downarrow\downarrow\downarrow\rangle)\nonumber\\
 &&+ \frac{5}{8}(\langle w|B|w\rangle+\langle\tilde{w}|B\tilde{w}\rangle), \\
 c_x(B)&=&-\frac{\sqrt{3}}{8}(\langle\uparrow\uparrow\uparrow\!|B|w\rangle+\langle w|B|\!\uparrow\uparrow\uparrow\rangle)\nonumber\\
 &&-\frac{\sqrt{3}}{8}(\langle\downarrow\downarrow\downarrow\!|B|\tilde{w}\rangle+\langle\tilde{w}|B|\!\downarrow\downarrow\downarrow\rangle)\nonumber\\
 &&-\frac{1}{4}(\langle w|B|\tilde{w}\rangle+\langle\tilde{w}|B|w\rangle),\\
 ic_y(B)&=&\frac{\sqrt{3}}{8}(\langle\uparrow\uparrow\uparrow\!|B|w\rangle-\langle w|B|\!\uparrow\uparrow\uparrow\rangle)\nonumber\\
 &&+\frac{\sqrt{3}}{8}(-\langle\downarrow\downarrow\downarrow\!|B|\tilde{w}\rangle+\langle\tilde{w}|B|\!\downarrow\downarrow\downarrow\rangle)\nonumber\\
 &&+\frac{1}{4}(\langle w|B|\tilde{w}\rangle-\langle\tilde{w}|B|w\rangle),\\
 c_z(B)&=&-\frac{3}{8}(\langle\uparrow\uparrow\uparrow\!|B|\!\uparrow\uparrow\uparrow\rangle-\langle\downarrow\downarrow\downarrow\!|B|\!\downarrow\downarrow\downarrow\rangle)\nonumber\\
 &&-\frac{1}{8}(\langle w|B|{w}\rangle-\langle\tilde{w}|B|\tilde{w}\rangle).\nonumber\\
 \end{eqnarray}
 
 Similar to the decorated honeycomb case, ${E^\rpf}^t(B)$ is  invariant in permuting $a$, $b$ and $c$ in the special form $B=a\otimes b\otimes c$, and this can be used to simplify some calculations.
 Let us use the lower-case $s$ to denote the spin-1/2 operators $s^u\equiv \sigma^u/2$ and recall that $\rho_1\equiv \openone/2$. One can then by direct calculation show that
 \begin{subequations}
 \label{eqn:Erpft}
 \begin{eqnarray}
 &&{E^\rpf}^t(\rho_1\otimes\rho_1\otimes\rho_1)=\frac{5}{8}\rho_1, \\
 &&{E^\rpf}^t(s^u\otimes s^u\otimes s^u)=-\frac{1}{8} s^u,\\
  &&{E^\rpf}^t(s^u\otimes s^v\otimes s^v)=-\frac{1}{24} s^u, \ {\rm for}\, u\ne v,\\
   &&{E^\rpf}^t(s^u\otimes s^v\otimes s^w)=0, \ {\rm for}\, u\ne v\ne w,\\
 &&{E^\rpf}^t(\rho_1\otimes s^u\otimes s^v)=\frac{5}{24}\delta_{uv}\rho_1,\\
 &&{E^\rpf}^t(\rho_1\otimes\rho_1 \otimes s^u)=-\frac{5}{24}s^u.
 \end{eqnarray}
 \end{subequations}
 To proceed further, it is useful to introduce 
 \begin{equation}
 \label{eqn:Lambda}
 \Lambda^u\equiv\openone \otimes s^u\otimes s^u+ s^u\otimes\openone\otimes s^u+s^u\otimes s^u\otimes\openone,
 \end{equation} and by direct calculation one can re-write Eq.~(\ref{eqn:EI}) as
 \begin{equation}
 \label{eqn:EI2}
 E^\rpf(\openone)=\frac{5}{4} \Pi^{S=3/2}_{\rm sym}=\frac{5}{8}(\openone\otimes\openone\otimes\openone+\frac{4}{3}\sum_{u=x,y,z} \Lambda^u),
 \end{equation}
which will allow us later to deduce $E^\rpf$ from the actions of $(E^\rpf)^t$ in Eqs.~(\ref{eqn:Erpft}) by fixing the overall scale.
 
 It is convenient to express the channel and its dual in the form of a matrix, sometimes called the superoperator form or the Liouville formalism. Thus, any matrices, such as $\sigma$, that the channels act on will be written in terms of vectors, such as $|\sigma\rangle\rangle$. Moreover the inner product between two such `vectors' becomes  $\langle\langle\sigma|\rho\rangle\rangle\equiv{\rm Tr}(\sigma^\dagger\rho)$.   Note that in this definition $\langle\langle\openone|\rho_1\rangle\rangle=1$.
 Then exploiting the permutation invariance of $E^{\rpf}$, one can employ the trick used in Ref.~\cite{Abdul-Rahman2019}, by using the action of the dual channel $(E^\rpf)^t$ in Eqs.~(\ref{eqn:Erpft}) and fixing the overall scale via Eq.~(\ref{eqn:EI2}), to deduce the action of $E^\rpf$ and write it in the `superoperator' form as
 \begin{eqnarray}
 {E^\rpf}&=&{5}|\rho_1\otimes\rho_1\otimes\rho_1\rangle\rangle\langle\langle \rho_1| -\sum_{u=x,y,z} |s^u s^u s^u\rangle\rangle \langle\langle s^u|\\
 &&+\frac{5}{3} \sum_{u}(|\rho_1 s^u s^u\rangle\rangle+| s^u \rho_1s^u\rangle\rangle+| s^u s^u\rho_1\rangle\rangle)\langle\langle \rho_1|\nonumber\\
 && -\frac{5}{3}\sum_u(|\rho_1\rho_1 s^u\rangle\rangle+| \rho_1 s^u \rho_1\rangle\rangle+| s^u \rho_1\rho_1\rangle\rangle)\langle\langle s^u|\nonumber\\
  &&\!\!\!\!\!\!-\frac{1}{3}\sum_u\sum_{v\ne u}(|s^v s^v s^u\rangle\rangle+| s^v s^u s^v\rangle\rangle+| s^u s^v s^v\rangle\rangle)\langle\langle s^u|,\nonumber
 \end{eqnarray}
 where we have suppressed the $\otimes$ symbols.
 It is also possible to calculate $E^\rpf$ directly from its definition in Eq.~(\ref{eqn:Erpf}), but the trick above helps to express $E^\rpf$ in terms of the sum of the product forms for the superoperators.
  
 From the results of Ref.~\cite{Abdul-Rahman2019}, the channel $E^n$ along $n$ decorated spin-1 sites,
 \begin{equation}
 \label{eqn:En}
 E^n(B)\equiv \sum_{i's} V_{i_n} ^\dagger.. V^\dagger_{i_1} B V_{i_1} .. V_{i_n} 
 \end{equation}
  is calculated to be
 \begin{equation}
 \label{eqn:E}
 E^n=|\openone\rangle\rangle\langle\langle\rho_1|+\frac{2(-1)^n}{3^n}\sum_u |s^u\rangle\rangle\langle\langle s^u|,
 \end{equation}
 and thus the combined channel from the left is (see Fig.~\ref{fig:tensors})
 \begin{eqnarray}
 \label{eqn:EL}
 E_L&=&(E^n\otimes E^n\otimes E^n\otimes) {E^\rpf}\\
 &=& \frac{5}{8}|\openone\openone\openone\rangle\rangle\langle\langle\rho_1| -\frac{(-1)^n}{3^{3n}}\sum_u(|s^u s^u s^u\rangle\rangle\langle\langle s^u|\\
 &&+\frac{5}{6\cdot 3^{2n}}\sum_u(|\openone s^u s^u\rangle\rangle+| s^u\openone s^u\rangle\rangle+|s^u s^u\openone\rangle\rangle)\langle\langle \rho_1|\nonumber\\
 &&-\frac{5(-1)^n}{12\cdot 3^n}\sum_u(|s^u\openone \openone\rangle\rangle+| \openone s^u\openone\rangle\rangle+|\openone\openone s^u \rangle\rangle)\langle\langle s^u|\nonumber \\
 &\!\!\! \!\!\!\!-&\!\!\!\!\frac{(-1)^n}{3^{3n-1}}\sum_u\sum_{v\ne u}(|s^v s^v s^u\rangle\rangle+| s^v s^u s^v\rangle\rangle+| s^u s^v s^v\rangle\rangle)\langle\langle s^u|\nonumber\\
  &=& \frac{5}{8}|\openone\openone\openone\rangle\rangle\langle\langle\rho_1| -\frac{(-1)^n}{ 3^{3n}}\sum_u(|s^u s^u s^u\rangle\rangle\langle\langle s^u|\\
 &&+\frac{5}{6\cdot 3^{2n}}\sum_u|\Lambda^u\rangle\rangle\langle\langle \rho_1|-\frac{5(-1)^n}{12\cdot 3^n}\sum_u|\Omega^u\rangle\rangle\langle\langle s^u|\nonumber\\
 &&-\frac{(-1)^n}{3^{3n-1}}\sum_u|\Theta^u\rangle\rangle\langle\langle s^u|,\nonumber
 \end{eqnarray}
 where $\Lambda^u$ was introduced in Eq.~(\ref{eqn:Lambda}) and here we introduce its vectorized form $|\Lambda^u\rangle\rangle\equiv |\openone s^u s^u\rangle\rangle+| s^u\openone s^u\rangle\rangle+|s^u s^u\openone\rangle\rangle$, as well as $|\Omega^u\rangle\rangle\equiv|s^u\openone \openone\rangle\rangle+| \openone s^u\openone\rangle\rangle+|\openone\openone s^u \rangle\rangle$ and $|\Theta^{u}\rangle\rangle\equiv\sum_{v\ne u} |s^v s^v s^u\rangle\rangle+| s^v s^u s^v\rangle\rangle+| s^u s^v s^v\rangle\rangle$.

Next, we consider the operator $Q_L\equiv E_L(\openone)$, and obtain it in the matrix form (instead of $|\cdots\rangle\rangle$)
 \begin{eqnarray}
 Q_L=\frac{5}{8}\openone\openone\openone+\frac{5}{6\cdot 3^{2n}}\sum_u \Lambda^u.
 \end{eqnarray}
 One can diagonalize $Q_L$ and obtain its spectrum (noting $\sum_u \Lambda^u$ has eigenvalues $\pm 3/4$)
 \begin{equation}
 {\rm spec}(Q_L)=\{\frac{5}{8} \pm\frac{5}{8\cdot 3^{2n}}\}.
 \end{equation}
 Therefore, the smallest eigenvalue $q_L$ of $Q_L$ is 
 \begin{equation}
 q_L=\frac{5}{8}-\frac{5}{8\cdot 3^{2n}}.
 \end{equation}
  The transfer operator  $E_L$ is completely positive~\cite{choi}, since it is constructed from Kraus operators via Eqs.~(\ref{eqn:Erpf}), (\ref{eqn:En}) and~(\ref{eqn:EL}), or alternatively it can be checked by directly diagonalizing the corresponding Choi matrix~\cite{choi}. Hence, it is also 2-positive, and 
 from a Cauchy-Schwarz inequality for 2-postive maps~\cite{Paulsen}, we have that
 \begin{eqnarray}
 \|E_L\|&=&\|E_L(\openone)\|\\
 &=&\|Q_L\|=\frac{5}{8}+\frac{5}{8\cdot 3^{2n}}.
 \end{eqnarray}
 From this we can calculate the associated $b_{L}(n)=8a(n)\|E_L\|/q_L$ and obtain
 \begin{equation}
 b_{L}(n)=\frac{8\cdot 3^{-n}(1+{3^{-2n}})}{1-{3^{-2n}}},
 \end{equation}
 where $a(n)=\big|\big|E^n-|\openone\rangle\rangle\langle\langle \rho_1|\big|\big|$ was previously obtained in Ref.~\cite{Abdul-Rahman2019} to be $3^{-n}$; but one can also calculate $a(n)$ directly from Eq.~(\ref{eqn:E}).
 
 \begin{figure}
 {\centering
 \includegraphics[width=0.4\textwidth]{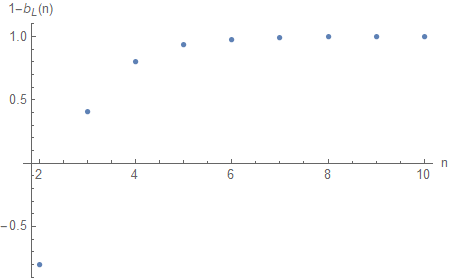}}
 \caption{\label{fig:oneminusbL}
 The function $1-b_L(n)$ vs. $n$, which is an indicator of injectivity the mapping $\Gamma_{L-C_n}$. Since $b_R(n)=b_L(n)$, this also applies to $\Gamma_{R-C_n}$.}
 \end{figure}
 \begin{figure}
 {\centering
 \includegraphics[width=0.4\textwidth]{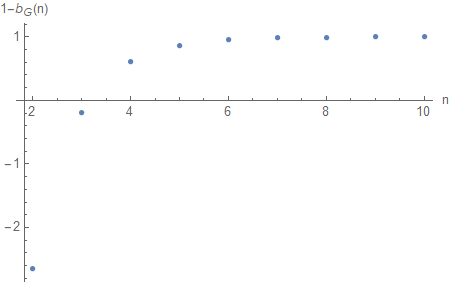}}
 \caption{\label{fig:oneminusbG}
 The function $1-b_G(n)$ vs. $n$, which is an indicator of injectivity for the mapping $\Gamma_{G}$.}
 \end{figure}
\begin{figure}
 {\centering
 \includegraphics[width=0.4\textwidth]{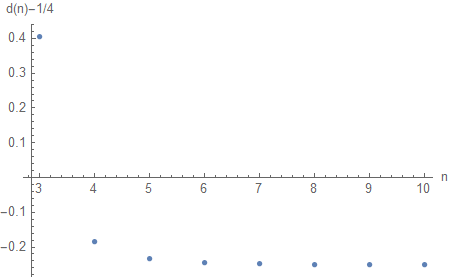}}
 \caption{\label{fig:epsilon_n}
 The function $d(n)-1/4$ vs. $n$. It is an indicator of a nonzero spectral gap if negative for the decorated square lattice.}
 \end{figure}

 Next, we examine the channel coming from the right square-lattice site. The onsite tensors are defined as $W_k^R\equiv 2\sqrt{2} (K\otimes K\otimes K) P_k^\dagger$,  and one finds that
 \begin{eqnarray}
 &&W_2^R=- (W_{-2}^L)^\dagger, \ W_{-2}^R=- (W_{2}^L)^\dagger, \\
 && W_1^R=(W_{-1}^L)^\dagger, \  W_{-1}^R=(W_{1}^L)^\dagger, \\
 && W_0^R=-(W_0^L)^\dagger. 
 \end{eqnarray} 
 From these, we see that $E^\lpf_R(B)\equiv\sum_k (W_k^R)^\dagger B W_k^R$ is dual to the channel $E^\rpf_L$, i.e., $E^\lpf_R=(E^\rpf_L)^{t}$. Therefore, using the superoperator formalism, $E_R\equiv E^\lpf\circ(E^n\otimes E^n\otimes E^n)$ is dual to $E_L$, i.e. $E_R=(E_L)^t$; this shows that $\|E_R\|=\|E_L\|$. Moreover, the operator $Q_R\equiv E_R^{t}(\rho_1)=E_L(\rho_1)=Q_L/2$, and therefore  we have that the relation between the minimum eigenvalues of $Q_R$ and $Q_L$ is $q_R=q_L/2$. We therefore obtain
 \begin{eqnarray}
 b_R(n)&=&4a(n) \|E_R\|/q_R=b_L(n),\\
 b_G(n)&=&4a(n)\|E_L\|\,\|E_R\|/(q_L q_R)\nonumber\\
 &=&8a(n) \|E_L\|^2/q_L^2.
 \end{eqnarray}
 The injectivity of the mappings $\Gamma_{G_{{L/R}-C_n}}$ and $\Gamma_G$ for the corresponding matrices $B$, $C$, $D$ to the respective quantum states, 
 \begin{eqnarray}
&& \Gamma_{G_{{L}-C_n}}(B)\equiv \sum_{l,i_1,..,i_n} {\rm Tr}
[B V_{i_n} .. V_{i_1} T_l^L] |l\rangle_L\otimes |i_1,..,i_n\rangle,\nonumber\\
&&\Gamma_{G_{{R}-C_n}}(C)\equiv \sum_{i_1,..,i_n,r} {\rm Tr}
[C T_r^R V_{i_n} .. V_{i_1} ]  |i_1,..,i_n\rangle\otimes|r\rangle_R,\nonumber\\
 &&\Gamma_{G}(D)\equiv
 \sum_{l,i's,r} {\rm Tr}
[D T_r^R V_{i_n}.. V_{i_1} T_l^L] |l\rangle_L\otimes |i_1,..,i_n\rangle\otimes|r\rangle_R,\nonumber
 \end{eqnarray}
  depends on whether $1-b_{L/R}(n)>0$ and $1-b_{G}(n)>0$, respectively; see Ref.~\cite{Abdul-Rahman2019}. 
  In the above equations, $T_l^L$ and $T_l^R$ denote tensors from the left and right sides, respectively, $|l\rangle_L$ and $|r\rangle_R$ are basis states for the left and right sides, respectively, and $V_i$ denotes the tensor for one spin-1 site that decorates the edge ($n$ is the total number of such sites); see Fig.~\ref{fig:tensors}. We have checked that $\Gamma_{G_{L-C_n}}$, $\Gamma_{G_{R-C_n}}$ and $\Gamma_G$ are injective for $n\ge 2$; see Figs.~\ref{fig:oneminusbL} and \ref{fig:oneminusbG}.  From Ref.~\cite{Abdul-Rahman2019}, $b_{LR}(n)\equiv b_L(n)+b_R(n)-b_L(n)b_R(n)=2b_L(n)-b_L(n)^2$, and it was shown that the important quantity $\varepsilon_n$  is upper bounded by
 \begin{equation}
 \varepsilon_n\le d(n)\equiv \frac{4a(n)}{\sqrt{1-b_{LR}(n)}}+ \left(\frac{4a(n)}{\sqrt{1-b_{LR}(n)}}\right)^2\big(1+b_G(n)\big).
 \end{equation}
 Here, if $d(n) <1/4$ then the corresponding AKLT model has a finite gap, whereas if  $d(n) >1/4$ it is undecided. 
 Thus, we have
 \begin{equation}
 d(n)= \frac{4a(n)}{|1-b_{L}(n)|}+ \left(\frac{4a(n)}{1-b_{L}(n)}\right)^2\big(1+b_G(n)\big).
 \end{equation}
 We can thus prove that the AKLT models on the decorated square lattice are gapped with $n\ge 4$, as shown in Fig.~\ref{fig:epsilon_n}. But the analytics cannot say anything about $n<4$.

	\begin{table}
	\begin{tabular}{|c | c | c |c| c|}
	\hline
 $n$ & \begin{tabular}{c}deg. 3, e.g. \\  honeycomb\end{tabular} & \begin{tabular}{c}deg. 4, e.g. \\  square\end{tabular} 
 & \begin{tabular}{c}mixed deg.  \\
  3\&4; Fig.~\ref{fig:lattices}i\end{tabular}& deg. 6 \\
	\hline
	1 & 0.4778328889 & 0.5234369088 & 0.5001917602& 0.6027622993 \\
	2 & 0.1183378500 & 0.1218467396 & 0.1200794787 &0.1285855428 \\
	3 & 0.0384373228 & 0.0389033280 & 0.0386700977& \\
	4 & 0.0124460198 & 0.0124961718 & 0.0124710706& \\
	5 & 0.0041321990 &  & & \\
	\hline
	\end{tabular}
	\caption{$\varepsilon_n$ for both the decorated honeycomb, square lattices and the lattice with mixed degrees 3 \& 4 (with 10 digits of accuracy presented). If $\varepsilon_n<1/3$ for the decorated honeycomb case, then we are sure that the corresponding AKLT model is gapped. For the decorated square lattice and the mixed-degree one, if $\varepsilon_n<1/4$, then we are sure that the corresponding AKLT model is gapped. For the decorated triangular lattice, if $\varepsilon_n<1/6$, then we are sure that the corresponding AKLT model is gapped.  From this table, we conclude that the AKLT models are gapped on all four types of decorated lattices with $n\ge2$.} \label{tbl:epsilon}
	\end{table} 
	
 Since $d(n)$ is only an upper bound on $\varepsilon_n$, we also performed numerical calculations directly for $\varepsilon_n$ for both the decorated honeycomb and square lattice (as well as one with mixed degrees), and confirmed that for both $n=2$ and $n=3$ the AKLT models are also gapped. The numerical results are shown in Table~\ref{tbl:epsilon}. We describe our methods in Sec.~\ref{sec:numerics}.
 
 \section{Comments on other lattice}
  \label{sec:honeycomb}

 \subsection{Other trivalent lattices}
 Since the proof in the decorated honeycomb case~\cite{Abdul-Rahman2019} only relies on the local structure of the two vertices on the underlying  lattice and the corresponding tensors  (see Fig.~\ref{fig:tensors} for illustration), a moment of thought will convince one that it also holds exactly for other trivalent lattices with decoration on their edges; see Fig.~\ref{fig:lattices} for illustration of other lattices. (However, this does not necessarily mean that the actual values of the gap will be identical.) Therefore for all trivalent lattices, which can be of any dimensions, such as 3D, the AKLT models on the corresponding decorated lattices will also be gapped if $n\ge 3$ (using the results on the decorated honeycomb in Ref.~\cite{Abdul-Rahman2019}), where again $n$ is the number of spin-1 sites added to decorate an edge.
In fact, for each undecorated edge, the number of decorated sites $n_e$ can be different, and the corresponding AKLT model will still be gapped as long as $n_e\ge 3$.  Numerically these bounds are improved to $n\ge 2$; see below.
\subsection{Other lattices of vertex degree 4}
By the same token, since we have proven that the AKLT models on the decorated square lattices are gapped if $n\ge 4$, this will also hold for any other decorated lattices, whose undecorated vertex degree is 4; see Fig.~\ref{fig:lattices}g\&h for illustration of such  lattices.    Numerically these bounds are also improved to $n\ge 2$.  AKLT states on the 3D diamond lattice (also four-valent) and the associated decorations are also universal~\cite{Wei2014,Wei2015}, and the significance is that these 3D resource states are likely to provide fault tolerance similar to the 3D cluster state~\cite{RHG}. Therefore, the decorated diamond lattices host AKLT models that are gapped for $n\ge2$, and the corresponding ground states  are also universal and likely provides topological protection for MBQC. 
\subsection{Other lattices of fixed vertex degree}
We conjecture that for any lattices of fixed vertex degree, the AKLT models on the corresponding decorated lattices will be gapped, as long as $n$ is large enough.  The intuition comes from that for large $n$, it is essentially many long spin-1 AKLT chains incident on some vertices, which act as local perturbations. For $n$ sufficiently large, the perturbation is of measure zero as $n\rightarrow \infty$.  Of course, this is only an intuition, rather than an actual proof.

 \begin{figure}
 {\centering
 \includegraphics[width=0.4\textwidth]{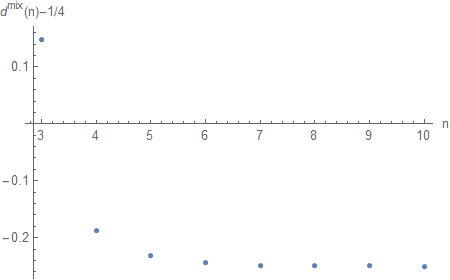}}
 \caption{\label{fig:epsilon_n_Mix}
 The function $d^{\rm mix}(n)-1/4$ vs. $n$. It is an indicator of a nonzero spectral gap if negative for a decorated lattice, whose underlying lattice has mixed degrees 3 and 4. One should evaluate $d^{\rm mix}(n)-1/4$ instead of $d^{\rm mix}(n)-1/3$ to check the gappedness.}
 \end{figure}
\subsection{Lattices of mixed vertex degrees}
A natural extension to examine is those AKLT models residing on decorated lattices whose undecorated ones are of mixed vertex degrees. It is likely that they will be gapped as long as $n$ is sufficiently large.

Let us consider the lattice (i) in Fig.~\ref{fig:lattices}, whose underlying lattice has mixed vertex degrees of 3 and 4. Take the left original site to be of degree 3 and the right original site of degree 4.  We have to evaluate $b_L(n)$, $b_R(n)$, $b_G(n)$, $E_L$ and $E_R$, and they can be obtained partly from the honeycomb case and partly from the square lattice case,
\begin{eqnarray}
&&b_L(n)=b_L^{\rm HC}(n)=\frac{8\cdot3^{-n} (1+3^{-2n-1})}{1-3^{-2n}},\\
&&b_R(n)=b_R^{\rm SQ}(n)=\frac{8\cdot 3^{-n}(1+ 3^{-2n})}{1-{3^{-2n}}},\\
&& b_{LR}(n)=b_L(n)+b_R(n)-b_L(n)b_R(n),\\
&& \|E_L\|=\|E_L^{\rm HC}\|=1+3^{-2n-1},\\
 &&\|E_R\|= \|E_R^{\rm SQ}\|=\frac{5}{8}(1+ 3^{-2n}),\\
 && b_G (n) =
8\cdot 3^{-n}
\|E_L\|\,\|E_R\| /(q_L^{\rm HC}q_R^{\rm SQ}),\\
&&q_L^{\rm HC}=1-3^{- 2n}, \ \ q_R^{\rm SQ}=\frac{5}{16}-\frac{5}{16\cdot 3^{2n}}.
\end{eqnarray}
Thus,  we obtain the corresponding function $d(n)$ for the mixed-degree lattice,
\begin{equation}
 d^{\rm mix}(n)= \frac{4\cdot 3^{-n}}{\sqrt{1-b_{LR}(n)}}+ \left(\frac{4\cdot 3^{-n}}{\sqrt{1-b_{LR}(n)}}\right)^2\big(1+b_G(n)\big).
 \end{equation}
 We see that the AKLT models are gapped for $n\ge 4$ for the decorated lattices, as checked in Fig.~\ref{fig:epsilon_n_Mix}.
Numerically these are improved to $n\ge 2$; see Table~\ref{tbl:epsilon}.
\section{Basis for Numerical Methods}
\label{sec:numerics}
Here we explain our numerical approach for producing the values of $\varepsilon_n$ in Table~\ref{tbl:epsilon}, which was derived  based on Lemma~6.3 of Ref.~\cite{Fannes1992}. The analytical results in the previous sections provide only upper bounds on $\varepsilon_n$, as inequalities such as operators norms and Schwarz inequalities were used in deriving, e.g., Eq.~(\ref{eqn:bound}). As we have seen, the analytics can only establish a nonzero gap for $n\ge 4$, but our numerical evaluation of $\varepsilon_n$ was able to push the gappedness to $n\ge 2$.

We begin by noting part (1) of the Lemma, which determines that
\begin{equation}
\varepsilon \equiv \|EF-E\wedge F\| =
\|(\mathbbm{1}-E)(\mathbbm{1}-F)-(\mathbbm{1}-E)\wedge(\mathbbm{1}-F)\|,
\end{equation}
where $E\wedge F$ projects onto the intersection of images
$E\mathcal{H} \cap F\mathcal{H}$ and, likewise, $E \vee F$ projects onto the
sum $E\mathcal{H} + F\mathcal{H}$, or
$(E\mathcal{H}^\perp\cap F\mathcal{H}^\perp)^\perp$.
From here on we will use $E\equiv \mathbbm{1}-P_v$ and
$F\equiv \mathbbm{1}-P_w$ rather than their complements, which will prove
useful because $P_v$ and $P_w$ are high-dimensional projectors and
their complements are low-dimensional.

Here we also review the findings that lead the source to part (2) of
the Lemma. In doing so, we will diverge from the source by \textit{not}
quotienting out $E\mathcal{H} \cap F\mathcal{H}$ and
$E\mathcal{H}^\perp \cap F\mathcal{H}^\perp$ (i.e. setting $E\wedge F=0$
and $E\vee F=\mathcal{1}$), as we ultimately will be working partly within
those spaces. We consider the eigenvalue equation
\begin{equation}
\label{eqn:EplusF}
(E+F)\Upsilon = (1-\alpha)\Upsilon.
\end{equation}
Clearly, as $E$ and $F$ are Hermitian operators whose eigenvalues belong to
$\{0,1\}$, the range of possible values of $\alpha$ is $[-1,+1]$. Moreover,
we note that $\alpha=-1$ corresponds exactly to the subspace
$E\mathcal{H}\cap F\mathcal{H}$, whereas $\alpha=+1$ corresponds exactly to the
subspace $E\mathcal{H}^\perp\cap F\mathcal{H}^\perp$. The remaining eigenspaces
must lie within the mutual orthogonal complement of these spaces,
\begin{align}
(E\mathcal{H}^\perp\cap &F\mathcal{H}^\perp)^\perp\cap(E\mathcal{H}\cap F\mathcal{H})^\perp \notag \\
&= (E\mathcal{H}+F\mathcal{H})\cap(E\mathcal{H}\cap F\mathcal{H})^\perp \notag \\
&= E\mathcal{H}\cap(E\mathcal{H}\cap F\mathcal{H})^\perp+F\mathcal{H}\cap(E\mathcal{H}\cap F\mathcal{H})^\perp \notag \\
&\equiv V_E+V_F,
\end{align}
noting that the explicit exclusion of $E\mathcal{H}\cap F\mathcal{H}$ from
$V_E$ and $V_F$ means that the above sum is a direct sum.

Therefore, for
$\alpha \neq \pm 1$, we can uniquely write $\Upsilon = \varphi+\psi$ for
$\varphi\in V_E$ and $\psi\in V_F$.
In particular we can rewrite Eq.~(\ref{eqn:EplusF}) as
\begin{align}
(E+F)(\varphi+\psi) &= (1-\alpha)(\varphi+\psi),
\end{align}
and arrive at
\begin{align}
(\varphi + E\psi) + (\psi + F\varphi) &= (\varphi - \alpha\varphi) + (\psi - \alpha\psi),
\end{align}
which we can rewrite as
\begin{equation}
(E\psi + \alpha\varphi) = -(F\varphi + \alpha\psi) \equiv \vartheta.
\label{eq:deftheta}
\end{equation}

We see immediately that $\vartheta \in E\mathcal{H}\cap F\mathcal{H}$.
Moreover, since we have constructed
$V_E,V_F\subset (E\mathcal{H}\cap F\mathcal{H})^\perp$, we immediately have
$\ipr{\vartheta}{\varphi}=\ipr{\vartheta}{\psi}=0$. Thus, we can for
example take $\bra{\vartheta}E\ket{\psi}$ and apply
$E$ to both the right and left,
\begin{align}
\bra{\vartheta}E\ket{\psi} &= -\alpha\ipr{\vartheta}{\psi}+\ipr{\vartheta}{\vartheta}=\|\vartheta\|^2\notag \\
&= \ipr{\vartheta}{\psi}=0.
\end{align}
That is, $\vartheta=0$, and consequently,
$E\psi = -\alpha\varphi$ and $F\varphi = -\alpha\psi$.

From this we can directly compute
\begin{equation}
(EF+FE)(\varphi+\psi) = -\alpha(1-\alpha)(\varphi+\psi)
\end{equation}
Such direct calculation also gives us
$EF+FE|_{E\mathcal{H}\cap F\mathcal{H}}=2$ and
$EF+FE|_{E\mathcal{H}^\perp\cap F\mathcal{H}^\perp}=0$. In particular,
consideration of individual eigenspaces gives us
\begin{equation}
EF+FE \geq -\max(\{\alpha\}\setminus\{1\})(E+F).
\end{equation}

We will then follow the original proof of part (1) of the Lemma in demonstrating
\begin{prop} The inequality
\begin{equation}
EF+FE \geq -\varepsilon(E+F)
\end{equation}
is optimized by $\varepsilon = \max(\{\alpha\}\setminus\{1\}) =
\|EF-E\wedge F\|$. In particular, $1-\varepsilon$ is the least nontrivial
eigenvalue of $E+F$.
\label{prop:EplusF}
\end{prop}

The operator norm $\|O\|$ is equivalent to the supremal real value of
$\langle \Phi|O|\Psi\rangle$ for unit $\Psi,\Phi$; in particular
optimizing $\Phi$ and $\Psi$ implies that
$O\Psi = \|O\|\Phi$ and $\|O\|\Psi = O^\dagger\Phi$.
In finding $\Psi$, we note that $EF-E\wedge F$ vanishes on both
$F\mathcal{H}^\perp$ and $E\mathcal{H}\cap F\mathcal{H}$; i.e. $\Psi$ is
orthogonal to these spaces and in particular $\Psi \in V_F$.
Likewise, the Hermitian transpose
vanishes on $E\mathcal{H}^\perp$ and $E\mathcal{H}\cap F\mathcal{H}$, so that
we should find $(EF-E\wedge F)\Psi \in V_E$; in particular, $\Phi \in V_E$.
Noting $(E\wedge F)\Psi=(E\wedge F)\Phi=0$, thus we can write $EF\Psi = E\Psi = \varepsilon\Phi$ and
$(EF)^\dagger\Phi=F\Phi=\varepsilon\Psi$. It follows that
\begin{equation}
(EF+FE)(\Psi-\Phi)=(\varepsilon^2-\varepsilon)(\Psi-\Phi)=-\varepsilon(E+F)(\Psi-\Phi)
\end{equation}

Moreover for \textit{any} eigenvector $\Upsilon$ of $E+F$ with eigenvalue
$1-\alpha\in(0,2)$, decomposed as above into $\varphi+\psi$,
$(EF-E\wedge F)\psi=EF\psi=-\alpha\varphi$. In particular,
this implies that $\|EF-E\wedge F\| \geq |\alpha|$, as $\psi$ and $\varphi$
have the same norm when $\alpha \neq 0$:
\begin{align}
\bra{\psi}E\ket{\varphi} &= -\alpha\ipr{\varphi}{\varphi} = \ipr{\psi}{\varphi} \notag \\
\bra{\psi}F\ket{\varphi} &= \ipr{\psi}{\varphi} = -\alpha\ipr{\psi}{\psi}
\end{align}
for $\alpha\neq \pm 1$; that is
$\varepsilon = \max(\{\alpha\}\setminus\{1\})$. $\square$

Therefore, determining $\varepsilon$ is equivalent to determining the
\textit{least nontrivial eigenvalue} of $E+F$. We now demonstrate that we can
simplify $E+F$ and, by extension, reduce the complexity of this calculation.

Consider a projector $A$, with the properties $EA=AE=E$ (i.e.
$A\mathcal{H}\supset E\mathcal{H}$) and $[A,F]=0$. (In particular, we will be
interested in a projector defined on the sites $Y_v\setminus Y_w$.)

\begin{prop}
For an eigenvector $\Upsilon$ of $E+F$ with eigenvalue $1-\alpha$,
$\alpha \notin \{-1,0,+1\}$, $A\Upsilon=\Upsilon$.
\label{prop:Aspectrum}
\end{prop}

As above, we write $\Upsilon=\varphi+\psi$ with $\varphi\in V_E$ and
$\psi\in V_F$, so that $E\psi=-\alpha\varphi$ and $F\varphi=-\alpha\psi$;
in particular $FE\psi=\alpha^2\psi$.
Manifestly $A\varphi=\varphi$ as $\varphi\in E\mathcal{H}$; meanwhile, since
$\alpha\neq 0$ we can write
\[ A\psi = \alpha^{-2}AFE\psi=\alpha^{-2}FAE\psi=\alpha^{-2}FE\psi=\psi.\square \]

We use $A$ to project onto a lower-dimensional subspace $\mathcal{H}'$;
that is we take $U_A:\mathcal{H}\to \mathcal{H}'$, for $U_A^\dagger U_A=A$
and $U_AU_A^\dagger=\mathbbm{1}_{\mathcal{H}'}$. We set $E'=U_AEU_A^\dagger$ and
$F'=U_AFU_A^\dagger$. That $E'$ and $F'$ are projectors follows directly
from the fact that $E$ and $F$ commute with $A$. Moreover,

\begin{prop} $\|E'F'+E'\wedge F'\|=\|EF+E\wedge F\|$
\end{prop}

We do this by examining the spectrum of $E'+F'$, as in
Prop.~\ref{prop:EplusF}. Since $A$ commutes with $E$ and $F$,
we find that
\begin{align*}
(E+F)U_A^\dagger \Upsilon' &= (E+F)U_A^\dagger (U_AU_A^\dagger)\Upsilon'\\
&= A(E+F)U_A^\dagger \Upsilon' = U_A^\dagger(E'+F')\Upsilon';
\end{align*}
that is, for any eigenvector $\Upsilon'$ of $E'+F'$, $U_A^\dagger\Upsilon'$
is an eigenvector of $E+F$ with the same eigenvalue. Put otherwise, the
spectrum  of $E'+F'$ is a subset of that of $E+F$. Then, by
Prop.~\ref{prop:Aspectrum}, only the degeneracies of eigenvalues
0, 1, and 2 are affected; in particular the least nontrivial eigenvalue
is preserved. $\square$

We additionally note that, for a fourth projector $B$ commuting with $E$ and
$A$ and satisfying $FB=BF=F$, $B'=U_ABU_A^\dagger$ satisfies the same
hypotheses for $F'$ and $E'$. Decomposing $B'=U_B^\dagger U_B$,
$U_BU_B^\dagger=\mathbbm{1}_{\mathcal{H}''}$, we can therefore move to a still smaller
space $\mathcal{H}'' \cong B'\mathcal{H}'$ and perform our analysis on
$E''=U_BE'U_B^\dagger$ and $F''=U_BF'U_B^\dagger$.
The method we use to efficiently exploit these conclusions is as
follows~\cite{footnote}:

\begin{enumerate}
  \item Determine $E=\mathbbm{1}-P_v$ as follows:
  \begin{enumerate}
    \item Construct the tensor corresponding to the portion of the AKLT state
    defined on $Y_v$, containing both physical and virtual indices 
    (in the honeycomb-lattice case, $3n+1$ physical and 3 virtual; in 
    the square-lattice case, $4n+1$ physical and 4 virtual indices).
    \item Collect the physical and virtual indices, in order to turn the
    representation into a matrix
    $\Psi \in \mathcal{H}_\text{phys}\otimes \mathcal{H}_\text{virt}$.
    \item Using the singular-value decomposition $\Psi=WsV^\dagger$ (written such that $s$ is full-rank), it follows
    that $E = WW^\dagger$.
  \end{enumerate}
  \item Taking $U_E=W^\dagger$, we can repeat this process to define
  isometries $U_F$ on $Y_w$, $U_A$ on $Y_v\setminus Y_w$, and
  $U_B$ on $Y_w\setminus Y_v$.
  \item Write $U_E'= U_E U_A^\dagger$ and $U_F'= U_F U_B^\dagger$ (as it may
  be prohibitively memory-intensive to represent even $E$ and $F$ in full).
  \item Then $E''=U_E^{\prime\dagger}U_E'$ and $F''=U_F^{\prime\dagger}U_F'$
  can be used to extract $\varepsilon$ by diagonalizing
  $E''+F''$.
\end{enumerate}

We applied the above procedure to four different types of lattices, and we found that the AKLT models are gapped for $n\ge 2$ for the decorated lattices, as shown in Table~\ref{tbl:epsilon}. This includes those whose underlying lattices are of degree 6, such as the triangular lattice and even the cubic lattice. The AKLT model on the cubic lattice is interesting, as the ground state, i.e. the AKLT state, is N\'eel ordered~\cite{Parameswaran}. By decorating the cubic lattice with a few spin-1 sites on every edge, the N\'eel order is removed, as gapless Goldstone modes must be present in the antiferromagnetic case. The results in Ref.~\cite{Wei2014} about quantum computational universality for the AKLT family only apply to lattices of vertex degrees equal to or less than 4. But for these 3D decorated AKLT states, we suspect that they are also universal for MBQC. 
\subsection{Lower bounds on the gap}
 The lower bound on the gap of the AKLTmodel on the decorated honeycomb can be estimated via Eq.~(\ref{eqn:upperlowerbounds}) and is given by
\begin{equation}
{\rm gap}(H_{\Lambda^{(n)}}^{\rm AKLT})\ge \frac{\gamma_Y(n)}{2}(1-3\varepsilon_n),
\end{equation}  
 shown in Ref.~\cite{Abdul-Rahman2019}.
The analytic bound of $\varepsilon_3<0.2683$ was used, and together with $\gamma_Y(n=3)\approx 0.2966$ this yielded
a lower bound of gap: ${\rm gap}(H_{\Lambda^{(n=3)}}^{\rm AKLT}) >0.0289$ for the decorated honeycomb lattice. Of course, this can be improved by using the numerical value for $\varepsilon_3$ from Table~\ref{tbl:epsilon}, and we obtain  ${\rm gap}(H_{\Lambda^{(n=3)}}^{\rm AKLT}) >0.131199$, which is four times more than orginally found. 

An additional improvement can be made by using a slightly different inequality from Eq.~(\ref{eqn:upperlowerbounds}): 
 \begin{equation}
 \label{eqn:upperlowerbounds2}
 {\Delta_Y} \tilde{H}_{\Lambda^{(n)}} \le H_{\Lambda^{(n)}}^{\rm AKLT}\le \|h'_{Y;v}\| \tilde{H}_{\Lambda^{(n)}},
 \end{equation}
 where $\Delta_Y(n)$ is defined 
 to be the smallest nonzero eigenvalue of $h'_{Y;v}$, which is similar to $h_v$ in Eq.~(\ref{eqn:HY}), but is instead defined as 
\begin{equation}
h'_{Y;v}= \sum_{e\in {\cal E}_{Y_v}\setminus {\cal E}_v} \frac{1}{2}P_e^{(z(e)/2)}+ \sum_{e\in {\cal E}_v} P_e^{(z(e)/2)},
\end{equation}
where ${\cal E}_v$ denotes the set of edges incident on the site $v$ on the original, undecorated lattice. The inequalites of Eq.~(\ref{eqn:upperlowerbounds2}) \revb{arise} naturally due to the fact that 
\begin{equation}
H_{\Lambda^{(n)}}^{\rm AKLT}=\sum_{v\in \Lambda} h'_{Y;v}.
\end{equation}
Thus the new lower bound on the gap is
\begin{equation}
{\rm gap}(H_{\Lambda^{(n)}}^{\rm AKLT})\ge  \gamma(n)\equiv\Delta_Y(n)(1-z\varepsilon_n),
\end{equation} 
where $z$ is the appropriate coordination number from the underlying lattice (one should take the largest one if the lattice is of mixed degree). We show in Table~\ref{tbl:gap} a few lower bounds on the gap. For the decorated honeycomb example  considered above, the lower bound on the gap is improved to
${\rm gap}(H_{\Lambda^{(n=3)}}^{\rm AKLT}) >0.183265$.

At this point, we would like to entertain the idea of extrapolating the lower bound from $n=3$ \& $n=2$ linearly to $n=1$ and $n=0$. Doing this, we would obtain ${\rm gap}(H_{\Lambda^{(n=1)}}^{\rm AKLT}) >0.1262096$ (extrapolated) and ${\rm gap}(H_{\Lambda^{(n=0)}}^{\rm AKLT}) >0.097682$ (extrapolated). The latter value is interesting, as it is consistent with the numerical gap value of the model on the honeycomb lattice $0.10$, obtained in Ref.~\cite{Garcia-Saez2013} using tensor-network methods. Of course, there is no basis for why such an extrapolation should be valid.
	\begin{table}
	\begin{tabular}{|c | c | c |c| c|}
	\hline
 $n$ & \begin{tabular}{c}$\Delta_Y(n)$   \\ for deg. 3 \end{tabular} & \begin{tabular}{c}gap lower \\ bound $\gamma(n)$   \end{tabular}& \begin{tabular}{c}$\Delta_Y(n)$ \\ for deg. 4 \end{tabular} & \begin{tabular}{c}gap lower \\ bound $\gamma(n)$   \end{tabular} \\
	\hline
	1 & 0.283484861 &  & 0.170646233&  \\
	2 & 0.239907874 & 0.154737328 & 0.197934811 &0.101463966 \\
	3 & 0.207152231 & 0.183265099 & & \\
	\hline
	\end{tabular}
	\caption{The local gap $\Delta_Y(n)$ for $h'_{Y;v}$ and the estimated lower bound on the gap $\gamma(n)$ for decorated AKLT models, whose underlying lattice, without decoration, has vertex degree 3 or 4. } \label{tbl:gap}
	\end{table}

 \section{Concluding remarks}
 \label{sec:conclude}
 We have followed the elegant approach by Abdul-Rahman et al.~\cite{Abdul-Rahman2019} and proved analytically that the decorated square lattices with $n\ge 4$ host AKLT models with finite spectral gap, similar to the results of the decorated honeycomb case. Our numerical approach extends beyond what was accessible previously and  allows to show that the AKLT models on both decorated lattices are gapped even for $n=2$ and $n=3$. The results of a nonzero spectral gap also hold for any other decorated lattices of which the underlying lattices are of fixed vertex degree 3 or 4. But we have also commented on other lattices. In particular, using the results from both the decorated honeycomb and square lattice, we also show analytically that AKLT models on decorated lattices where the underlying lattice has mixed vertex degrees 3 and 4, are also gapped for $n\ge 4$. This is improved numerically to $n\ge 2$. Regarding the spectral gap for the AKLT models on the undecorated honeycomb or square lattice, we also share the same view as the authors of Ref.~\cite{Abdul-Rahman2019}, i.e.  to establish their spectral gap will require a different and maybe novel approach.  However, some insight may be obtained if one can make progress analytically on the cases of $n=1,2$ and in particular whether $n=1$ case is gapped or not, for which we strongly suspect that it is gapped. 
 
 Our numerical results also show the nonzero gap for $n=2$ in the decorated triangular and cubic lattices.  Observing the decaying trend of $\varepsilon_n$  on $n$ in the previous analysis, we believe that the nonzero gap should exist for all $n\ge 2$.  One can carry out the analytic procedure for the degree-6 case. The calculations are expected to be more tedious but likely straightforward. Such a result is interesting for the cubic lattice, as this shows the AKLT states on the decorated cubic lattices are not N\'eel ordered, in contrast to the state on the undecorated cubic lattice. Naively, decoration using spin-1 sites introduces more quantum fluctuations than those from spin-3 sites and destroys the N\'eel order. In contrast, the ground state of the spin-1/2 Heisenberg model  on the cubic lattice is antiferromagnetically ordered, despite the seemingly larger quantum fluctuations from such low-spin magntude entities.   The phenomena of the suppresion of order, as well as the other kind of suppression---of frustration, as mentioned in the Introduction, may be of interest for futher exploration.
 
 AKLT models that have spin rotational symmetry but a deformation that breaks the full SO(3) symmetry were considered, such as the  deformed AKLT models in Refs.~\cite{Niggemann1997,Niggemann2000}. Can we employ a similar approach to prove the spectral gap for the deformed models on the decorated  lattices? It is also possible that ideas from tensor network can be useful, such as those in Refs.~\cite{Schuch2011,Darmawan2016}. Some deformed AKLT states were also previously shown to provide a universal resource for MBQC within some finite range of deformation~\cite{Darmawan2012,Huang2016}. These deformed models also have interesting phase diagrams~\cite{Niggemann1997,Niggemann2000,Hieida1999,Huang2016,Pomata2018}.
It is worth mentioning that some related 2D Hamiltonians interpolating the AKLT  and the cluster-state models were also shown to have finite spectral gap~\cite{Darmawan2016}, but the spectral gap in the exact AKLT limit is still not proved.

\begin{acknowledgments}
This work was supported by the National Science Foundation under grants No. PHY  1620252 and No. PHY 1915165. T.-C.W. thanks Bruno Nachtergaele for useful communication regarding the issue of the norms.
\end{acknowledgments}


\begin{thebibliography}{99}
\bibitem{footnote}Note that we have made extensive use of the implied
  identity tensors common in quantum informational notation. For example,
  we may for completeness write
  $U_E'=U_E(U_A^\dagger\otimes\mathbbm{1}_{Y_v\cap Y_w})$ and
  $E'' = (U_E^{\prime\dagger}U_E')\otimes\mathbbm{1}_B$.
\end{thebibliography}
\end{document}